\documentclass[showpacs, oneside, twocolumn, prd, amsmath, amssymb, nofootinbib, superscriptaddress]{revtex4-1}

\usepackage{cases}
\usepackage{amsmath}
\usepackage{amssymb}
\usepackage{amsfonts}
\usepackage{amssymb}
\usepackage{dcolumn}
\usepackage{bm}
\usepackage{bbm}
\usepackage{graphicx}
\usepackage{xcolor}
\usepackage{array}
\usepackage{subfigure}
\usepackage{hyperref}
\usepackage{wasysym}
\usepackage{mathrsfs}

\newcommand{\be}{\begin{equation}}
\newcommand{\ee}{\end{equation}}
\newcommand{\ba}{\begin{eqnarray}}
\newcommand{\ea}{\end{eqnarray}}

\newcommand{\gsim}{\mathrel{\hbox{\rlap{\lower.55ex \hbox {$\sim$}}
                   \kern-.3em \raise.4ex \hbox{$>$}}}}
\newcommand{\lsim}{\mathrel{\hbox{\rlap{\lower.55ex \hbox {$\sim$}}
                   \kern-.3em \raise.4ex \hbox{$<$}}}}

\hypersetup{colorlinks=true,
            breaklinks=true,
            pdfstartview=Fit,
            linkcolor=blue,
            citecolor=blue,
            urlcolor=blue}

\begin{document}
\title{New test on the Einstein equivalence principle through the photon ring of black holes}

\author{Chunlong Li}
\email{chunlong@mail.ustc.edu.cn}
\affiliation{Department of Astronomy, School of Physical Sciences, University of Science and Technology of China, Hefei, Anhui 230026, China}
\affiliation{CAS Key Laboratory for Research in Galaxies and Cosmology, University of Science and Technology of China, Hefei, Anhui 230026, China}
\affiliation{School of Astronomy and Space Science, University of Science and Technology of China, Hefei, Anhui 230026, China}

\author{Hongsheng Zhao}
\email{hz4@st-andrews.ac.uk}
\affiliation{Department of Astronomy, School of Physical Sciences, University of Science and Technology of China, Hefei, Anhui 230026, China}
\affiliation{Scottish Universities Physics Alliance, University of St Andrews, North Haugh, St Andrews, Fife KY16 9SS, UK}

\author{Yi-Fu Cai}
\email{yifucai@ustc.edu.cn}
\affiliation{Department of Astronomy, School of Physical Sciences, University of Science and Technology of China, Hefei, Anhui 230026, China}
\affiliation{CAS Key Laboratory for Research in Galaxies and Cosmology, University of Science and Technology of China, Hefei, Anhui 230026, China}
\affiliation{School of Astronomy and Space Science, University of Science and Technology of China, Hefei, Anhui 230026, China}

\begin{abstract}
Einstein equivalence principle (EEP), as one of the foundations of general relativity, is a fundamental test of gravity theories. In this paper, we propose a new method to test the EEP of electromagnetic interactions through observations of black hole photon rings, which naturally extends the scale of Newtonian and post-Newtoian gravity where the EEP violation through a variable fine structure constant has been well constrained to that of stronger gravity. We start from a general form of Lagrangian that violates EEP, where a specific EEP violation model could be regarded as one of the cases of this Lagrangian. Within the geometrical optical approximation, we find that the dispersion relation of photons is modified: for photons moving in circular orbit, the dispersion relation simplifies, and behaves such that photons with different linear polarizations perceive different gravitational potentials. This makes the size of black hole photon ring depend on polarization. Further assuming that the EEP violation is small, we derive an approximate analytic expression for spherical black holes showing that the change in size of the photon ring is proportional to the violation parameters. We also discuss several cases of this analytic expression for specific models. Finally, we explore the effects of black hole rotation and derive a modified proportionality relation between the change in size of photon ring and the violation parameters. The numerical and analytic results show that the influence of black hole rotation on the constraints of EEP violation is relatively weak for small magnitude of EEP violation and small rotation speed of black holes.
\end{abstract}



\maketitle

\section{Introduction}
\label{introduction}

The Event Horizon Telescope (EHT) has captured the first image of a supermassive black hole \cite{Akiyama:2019cqa, Akiyama:2019brx, Akiyama:2019sww, Akiyama:2019bqs, Akiyama:2019fyp, Akiyama:2019eap}. This provides us possibilities of probing new physics in a strong gravitational field. In the center of the galaxy M87, the compact radio source is resolved as an asymmetric bright emission disk, which encompasses a central dark region. In the current literature, although the details remain to be examined, it is pointed out that there could exit a strong lensing structure which is called ``photon ring" behind the dominated direct emission profile from the accretion disk \cite{Gralla:2019xty, Narayan:2019imo}. With the help of sufficient high-resolution imaging, complexities from astrophysical effects can be mitigated, as the size and shape of the photon ring are totally determined by the instabilities of photon orbits predicted by geodesic equations, which makes black hole photon ring become a potential probe to test gravity and related new physics \cite{Johnson:2019ljv, Gralla:2020nwp, Gralla:2020yvo, Gralla:2020srx, Gralla:2020pra}.

The high spatial resolution image taken by the EHT and the great potential of the photon ring on testing gravity have inspired a series of work. One type of the works focuses on the possible contamination of the perfect vacuum environments around the black hole, which could arise from the coupling of gravity to other background fields and the accumulation of dark matter due to accretion. Such novel physics modify the black hole metric and leave observable effects on the black hole photon ring \cite{Banerjee:2019xds, Pantig:2020uhp, Konoplya:2019sns, Jusufi:2020cpn, Jusufi:2019nrn, Chen:2020qyp, Huang:2016qnl}. On the other side, the question is about testing gravity theories. Some modified gravity theories could have different black hole solutions from those of the general relativity and thus lead to different patterns of photon motion \cite{Held:2019xde, Kumar:2019ohr, Cai:2010zh, Cai:2012db, Cai:2015fia, Zhu:2019ura, Amarilla:2015pgp, Ovgun:2019jdo, Ovgun:2020gjz, Wei:2020ght, Kumar:2020hgm, Tian:2019yhn, Guo:2020zmf, Vagnozzi:2019apd, Khodadi:2020jij, Long:2019nox, Banerjee:2019nnj}. We refer readers to \cite{Psaltis:2020lvx} for constraints on gravity theories under a parameterized post-Newtonian formalism. For these two types of works, in addition to mass, spin and electric charge, more parameters are introduced to describe the spacetime around black holes. These lines of works can thus be effectively regraded as theories that violate the no-hair theorem of black holes.

Besides violation of no-hair theorem, another manifestation of new physics beyond general relativity is the breakdown of Einstein equivalence principle (EEP). In standard general relativity, the unique gravitational field described by the metric is minimally coupled to the cosmic components including matter and interactions. This means gravity plays the role of a geometrical background: in a local free-falling frame where geometrical effects are canceled by the local transformations of reference frames, the fundamental non-gravitational physics return to those without gravity \cite{Will:2014kxa,Tino:2020nla,DiCasola:2013iia}. If the cosmic components are non-minimally coupled to gravity, or coupled to other unknown background fields, the coupling effects generally cannot be canceled by local transformations of the reference frame, and thus the EEP is not valid anymore. Therefore, whether EEP is established or not contains the information about the coupling among different components of the Universe to gravity, which allowing us to test possible new physics. We refer readers to \cite{Tino:2020nla} for a more precise introduction of the EEP and other kinds of equivalence principle.

One of the main challenges of modern physics is the confirmation of the EEP at different scales and different contexts both from the theoretical and the experimental points of view \cite{Wetterich:2002ic,Peccei:1987mm,Hui:2009kc,Kraiselburd:2015vyf,Landau:2001qb,Ni:1977zz,Donoghue:1984zs,Donoghue:1984ga,Kostelecky:2010ze,Hohensee:2011wt}. However, the current experiments on testing EEP are mainly conducted on the scale of Newtonian or post-Newtonian gravity \cite{Angelil:2011cq,Amorim:2019xrp,Hees:2020gda,Hees:2015nlf,Sandvik:2001rv,Bekenstein:1982eu,Wei:2018ajw,Giani:2020fpz,Possel:2019zbu,Bertolami:2017opd}. As for more extreme gravitational field such as that around the horizon of black holes, whether EEP holds or not is still unknown.

Black hole photon ring provides us with the opportunity to test EEP in the extremely strong gravitational filed. It is the lensed image of unstable bound orbits of photons around black hole. For Schwarzschild black hole, these orbits are circular with their radius equals to only three times the gravitational radius. Therefore, black hole photon ring, as the observable of unstable bound orbits, could become a potential probe to detect new physics in the gravitational field near the horizon scale. 

Possibility of violation of the EEP near a black hole could be suggested by the superradiance process of rotating black holes \cite{Roy:2019esk,Bar:2019pnz,Davoudiasl:2019nlo,Cunha:2019ikd,Chen:2019fsq}, which implies there might be a fruitful environment of light particles around rotating black holes at horizon scale \cite{ZelDovich:1972,Detweiler:1980uk,Nielsen:2019izz,Baumann:2019eav,Huang:2019xbu,Pawl:2004bx,Arvanitaki:2014wva}. Moreover, if these particles are those beyond the standard model and are coupled to photons, there might be a phenomenological violation of EEP. Furthermore, when we consider the effects of quantum field theory in curved spacetime, the vacuum polarization of photons can introduce a non-minimal coupling of electromagnetic fields to the spacetime curvature \cite{Berends:1975ah,Drummond:1979pp,Milton:1977je,Cai:1998ij}. This could also lead to a violation of EEP.

In this paper, we put forward a new method to test the EEP by using the observations of black hole photon ring. In Sec. II, we focus on the EEP violation occurring in electromagnetic interactions and start from a general Lagrangian that describes a new background field coupled with the electromagnetic field. In Sec. III, we apply the geometric optics approximation to derive a modified dispersion relation of photons and obtain the corresponding phenomenological behavior by restricting our discussions to a system with the static and spherical symmetry. Then in Sec. IV and Sec. V, we derive and show a connection between the size of the photon ring and the parameters of the EEP violation. We also give several specific examples of the EEP violation in Sec. VI. Finally, in Sec. VII, we generalize our discussions to rotating black holes and study the influence of all kinds of black hole parameters on the constraints of violation parameters. Sec. VIII is a summary of the main results with a discussion and a future outlook. We work in units where the gravitational constant $G=1$ and the speed of light $c=1$ and we adopt the metric convention $(-,+,+,+)$.

\section{The model of EEP violation}

There are several characteristic scales for an electromagnetic system in a curved spacetime. One is the varying scale $\lambda$ of the electromagnetic field, the other is the characteristic length $L_R$ of the spacetime curvature. And if there exists some additional background fields that are coupled to this electro-gravitational system, more length scales $L_Q$ characterizing these fields would be involved. The geometric optics approximation states that if $\lambda$ is much smaller than another other characteristic scales of the system, i.e. $\lambda\ll \min\{L_R,L_Q\}$, the electromagnetic vector $A_{\mu}$ generally have the following solution \cite{Misner}
\begin{align}
    A_{\mu}(x)=a_{\mu}(\epsilon,x)e^{\frac{i}{\epsilon}S(x)},
	\label{goa}
\end{align}
where $\epsilon$ is a small quantity which represents a rapidly varying phase and the expression of $a_{\mu}$ is
\begin{align}
	a_{\mu}(\epsilon,x)=\sum_{m=0}^{\infty}\left(\frac{\epsilon}{i}\right)^m a_m(x).
\end{align}
At the lowest order of the geometric optics approximation, i.e. the order of $1/\epsilon^2$, the wave equation gives rise to the dispersion relation of the 4-wave vector $k_{\mu}=(1/\epsilon)\partial_\mu S$, which determines the equation of motion of test photons. Intuitively, we could generally write the dispersion relation as
\begin{align}
	b_{\mu}k^{\mu}+c_{\mu\nu}k^{\mu}k^{\nu}+\mathcal{O}(k^3)=0,
	\label{gdr}
\end{align}
where $b_{\mu}$ and $c_{\mu\nu}$ are vector and tensor fields. $\mathcal{O}(k^3)$ represents the potential correction terms that are higher than the second power of $k^{\mu}$. In the current literature, there exists the following three cases for $\mathcal{O}(k^3)=0$:
\begin{itemize}
	\item $b_{\mu}=0$, $c_{\mu\nu}=g_{\mu\nu}$. This corresponds to the standard case $g_{\mu\nu}k^{\mu}k^{\nu}=0$ in the general relativity, where $g_{\mu\nu}$ is spacetime metric.
	\item $b_{\mu}\neq 0$, $c_{\mu\nu}=g_{\mu\nu}$. This case describes a background vector field couples with the motion of photons \cite{Li:2019lsm}. An example is the correction to the lowest order geometric approximation due to the large spacetime curvature, where $b_{\mu}$ is constructed by the spacetime curvature tensor and the null tetrad of photons \cite{Frolov:2020uhn,Frolov:2012zn}.
	\item $b_{\mu}=0$, $c_{\mu\nu}\neq g_{\mu\nu}$. A background field non-minimally coupled to the electromagnetic tensor often leads to this kind of modification. An example is the non-minimal coupling of electromagnetic field to spacetime curvature induced by the virtual electron loops in quantum electrodynamics \cite{Berends:1975ah,Drummond:1979pp,Milton:1977je,Cai:1998ij}, which could give a $c_{\mu\nu}$ different from the metric $g_{\mu\nu}$.
\end{itemize}

A modified dispersion relation given by Eq. (\ref{gdr}) often manifests itself by violation of the EEP. In this paper, we focus on the third cases, i.e. the quadratic correction $c_{\mu\nu}$. Firstly, let us consider the below electromagnetic Lagrangian which preserves the diffeomorphism and $U(1)$ gauge invariance
\begin{align}
	\mathcal{L}_{em}=-\frac14F_{\mu\nu}F^{\mu\nu}-\frac{1}{8}q Q^{\mu\nu\rho\sigma}(x)F_{\mu\nu}F_{\rho\sigma}-eJ^{\mu}A_{\mu},
	\label{La}
\end{align}
where $F_{\mu\nu}=\nabla_{\mu}A_{\nu}-\nabla_{\nu}A_{\mu}$ is the electromagnetic tensor and $\nabla_{\mu}$ is the covariant derivative with respect to the Levi-Civta connection. The final term describes the coupling to the matter current $J^{\mu}$ through the charge $e$ and we does not written down the corresponding matter action since it is not related to our discussions. The second term is beyond the standard physics, which describes an unknown background field $Q^{\mu\nu\rho\sigma}$ is non-minimally coupled to electromagnetic tensor through the coupling constant $q$. Because of the index symmetry of $F_{\mu\nu}F_{\rho\sigma}$, the field $Q^{\mu\nu\rho\sigma}$ should satisfy $Q^{[\mu\nu]\rho\sigma}=Q^{\mu\nu[\rho\sigma]}=Q^{\mu\nu\rho\sigma}$ and the exchanging symmetry of $\mu\nu$, $\rho\sigma$ as a whole.

In principle, $Q^{\mu\nu\rho\sigma}(x)$ could be a scalar, vector, tensor or a sum of these parts. The current work on testing action (\ref{La}) mainly focus on the scalar part of $Q^{\mu\nu\rho\sigma}(x)$. For example, if $Q^{\mu\nu\rho\sigma}$ is a scalar field $\phi$, i.e. $Q^{\mu\nu\rho\sigma}=2f(\phi)g^{\rho[\mu}g^{\nu]\sigma}$\cite{Sandvik:2001rv,Bekenstein:1982eu}, varying the Lagrangian (\ref{La}) with respect to $A_{\mu}$, one will obtain the modified Maxwell equations 
\begin{align}
	\nabla_{\mu}\left[\left(1-q f(\phi)\right)F^{\mu\nu}\right]=eJ^{\nu}.
\end{align}
For the practical case, $\phi$ should only vary little over large distances and times. Thus, $q f(\phi)$ could be taken out of the derivative, which is equivalent to replacing the electric charge $e$ with a field $e'=e/(1-q f(\phi))$. This fact tells us that the fine structure constant will have a variable value over the spacetime, which is often called the violation of the local position invariance (LPI) in the literature, as one of the elements of the EEP \cite{Will:2014kxa,Tino:2020nla,DiCasola:2013iia}. Then for the experimental test of this new coupling, one could detect whether atomic spectra at different locations represent the same fine structure constant such as a given kind of atom on Earth and the same kind of atom on stars in orbits of supermassive black holes. We refer readers to \cite{Angelil:2011cq,Amorim:2019xrp,Hees:2020gda} for more details.

As for the vector or tensor parts of $Q^{\mu\nu\rho\sigma}(x)$, the above method could not give rise to a simple result characterizing by a varying fine structure constant. Another defect of the above method is that it is an indirect test on the EEP violation Lagrangian (\ref{La}), which depends on the coupling to the matter field, i.e. $-eJ^{\mu}A_{\mu}$ and might not exclude the influence of properties of matter itself. In order to explore whether there exists the EEP violation term in the Lagrangian (\ref{La}), one need to seek a direct method to test this term.

\section{The phenomenological behavior of the EEP violation model}

In this section, we apply the geometric optics approximation to the first and second term of the Lagrangian (\ref{La}). This leads to a modified dispersion relation $c_{\mu\nu}k^{\mu}k^{\nu}=0$ under some conditions where $c_{\mu\nu}$ is no longer the spacetime metric and depends on the polarizations of photons. Therefore, photons with different polarizations could follow different propagation paths, which thus violates the weak equivalence principle (WEP), as another element of the EEP \cite{Will:2014kxa,Tino:2020nla,DiCasola:2013iia}. In the following, we will explain how this mechanism works.

Let us neglect the last term in the Lagrangian (\ref{La}) and vary this action with respect to 4-vector potential $A_{\mu}$, which gives
\begin{align}
	\nabla_{\mu}\left(F^{\mu\nu}-\frac12q Q^{\mu\nu\sigma\rho}F_{\rho\sigma}\right)=0.
	\label{ME}
\end{align}
After applying the geometric approximation (\ref{goa}) and only retaining the lowest order terms, the combination of the Lorentz gauge $\nabla_{\mu}A^{\mu}=0$ and the modified Maxwell equation (\ref{ME}) gives rise to
\begin{align}
	& k_{\sigma}k_{\mu}\left(q Q^{\mu\nu\sigma\rho}+g^{\sigma\mu}g^{\rho\nu}\right)a_{\rho}=0, 
	\label{dr}
\end{align}
Eq. (\ref{dr}) implies a modified dispersion relation of photons, which contains the information about the path of photons in spacetime. When $q=0$, this equation gives rise to the null curve $k^{\mu}k_{\mu}=0$ of the motion of photons in the standard general relativity. The Lorentz gauge $\nabla_{\mu}A^{\mu}=0$ gives rise to $k_{\mu}a^{\mu}=0$ under the geometric optics approximation. In order to take advantage of this feature to simplify the Eq. (\ref{dr}), one could introduce the antisymmetry basis \cite{Drummond:1979pp}
\begin{align}
	U^{\mu\nu}_{ab}=e^{\mu}_ae^{\nu}_b-e^{\nu}_ae^{\mu}_b,
\end{align}
where $e^{\mu}_a$ are the tetrad fields with $a=0, 1, 2, 3$, which satisfy $g_{\mu\nu}=\eta_{ab}e^a_{\mu}e^b_{\nu}$. The tetrad indices $a, b...$ are raised and lowered by $\eta_{ab}$. In this paper, given the index symmetry of tensor $Q^{\mu\nu\sigma\rho}$, we consider $Q^{\mu\nu\sigma\rho}$ has the below expansion form
\begin{align}
	Q^{\mu\nu\sigma\rho}&=C^{0101}(x)U^{\mu\nu}_{01}U^{\sigma\rho}_{01}+C^{0202}(x)U^{\mu\nu}_{02}U^{\sigma\rho}_{02} \nonumber \\
	&+C^{0303}(x)U^{\mu\nu}_{03}U^{\sigma\rho}_{03}+C^{1212}(x)U^{\mu\nu}_{12}U^{\sigma\rho}_{12} \nonumber \\
	&+C^{1313}(x)U^{\mu\nu}_{13}U^{\sigma\rho}_{13}+C^{2323}(x)U^{\mu\nu}_{23}U^{\sigma\rho}_{23},
	\label{expansion}
\end{align}
where the expansion coefficients $C^{abcd}(x)$ are functions of the spacetime coordinate $x$. By introducing the projection of $k^{\mu}$ on the antisymmetry basis
\begin{align}
	l_{\nu}\equiv k^{\mu}U^{01}_{\mu\nu}, \\
	m_{\nu}\equiv k^{\mu}U^{02}_{\mu\nu}, \\
	n_{\nu}\equiv k^{\mu}U^{03}_{\mu\nu},
\end{align}
together with the independent projections
\begin{align}
	p_{\nu}&\equiv k^{\mu}U^{12}_{\mu\nu}=\frac{1}{k^0}\left(k^1m_{\nu}-k^2l_{\nu}\right), \\
	q_{\nu}&\equiv k^{\mu}U^{13}_{\mu\nu}=\frac{1}{k^0}\left(k^1n_{\nu}-k^3l_{\nu}\right), \\
	r_{\nu}&\equiv k^{\mu}U^{23}_{\mu\nu}=\frac{1}{k^0}\left(k^2n_{\nu}-k^3m_{\nu}\right),
\end{align}
the dispersion relation (\ref{dr}) could be written as
\begin{equation}
	\begin{pmatrix}
	K_{11} & K_{12} & K_{13} \\
	K_{21} & K_{22} & K_{23} \\
	K_{31} & K_{32} & K_{33} \\
\end{pmatrix}
\begin{pmatrix}
	a\cdot l \\
	a\cdot m \\
	a\cdot n
\end{pmatrix}
=0,
\label{system}
\end{equation}
where the expressions of the matrix components are
\begin{align}
	K_{11}&=k\cdot k+q C^{0101}(k^0k^0-k^1k^1)+q C^{1212}k^2k^2 \nonumber \\
	&+q C^{1313}k^3k^3, \\
	K_{12}&=-q C^{0202}k^2k^1-q C^{1212}k^2k^1, \\
	K_{13}&=-q C^{0303}k^3k^1-q C^{1313}k^3k^1, \\
	K_{21}&=-q C^{0101}k^2k^1-q C^{1212}k^2k^1, \\
	K_{22}&=k\cdot k+q C^{0202}(k^0k^0-k^2k^2)+q C^{1212}k^1k^1 \nonumber \\
	&+q C^{2323}k^3k^3, \\
	K_{23}&=-q C^{0303}k^3k^2-q C^{2323}k^3k^2, \\
	K_{31}&=-q C^{0101}k^3k^1-q C^{1313}k^3k^1, \\
	K_{32}&=-q C^{0202}k^3k^2-q C^{2323}k^3k^2, \\
	K_{33}&=k\cdot k+q C^{0303}(k^0k^0-k^3k^3)+q C^{1313}k^1k^1 \nonumber \\
	&+q C^{2323}k^2k^2.
\end{align}

In order to simplify the system (\ref{system}), one could diagonalize this matrix and the condition that the product of the eigenvalues (the determinant of the matrix) equals to zero will give rise to the criterion of non-zero solutions of this system, which thus implies the dispersion relation of photons' motion. However, from the analysis of the actual physical situation, we could make several assumptions on this system firstly.

We focus on the static spherical spacetime and the metric is
\begin{align}
	ds^2=g_{tt}(u)dt^2+g_{rr}(u)dr^2+r^2(d\theta^2+\sin^2\theta d\phi^2),
\end{align}
where $u=M/r$ and $M$ is the mass of the central black hole. The asymptotically flat condition requires $g_{tt}(0)=-1$ and $g_{rr}(0)=1$ in the limit of far distance from the black hole. The corresponding tetrad fields could be written as
\begin{align}
	e^a_{\mu}=diag\left(\sqrt{-g_{tt}(r)}, \sqrt{g_{rr}(r)}, r, r\sin\theta\right).
\end{align}

Now we make two assumptions. The first one is that the motion of photons follow the same symmetry as the spacetime. Therefore, there should exist the integral constant of motion making the orbits of photons be bound in a plane, i.e. $\theta=\pi/2$ and $k^2=k^{\theta}e^2_{\theta}=0$. And this assumption also accounts for the reason why we focus on the expansion (\ref{expansion}) without the cross terms of $U^{\mu\nu}_{ab}$ such as $C^{0102}(x)U^{\mu\nu}_{01}U^{\sigma\rho}_{02}$ since these terms will break this assumption. The second assumption is that there exits circular orbits in the motion plane which satisfy $k^1=k^re^1_r=0$. The reason of adopting this assumption is that in the numerical simulation, the presence of the photon ring is caused by the existence of the unstable bound circular orbits, which makes it observationally interesting \cite{Gralla:2019xty}. Then under these two assumptions, the only non-zero matrix components for the photons moving in a plane circular orbits are
\begin{align}
	K_{11}&=(1-q C^{0101})g_{tt}k^tk^t+(1+q C^{1313})r^2k^{\phi}k^{\phi}, \label{K11} \\
	K_{22}&=(1-q C^{0202})g_{tt}k^tk^t+(1+q C^{2323})r^2k^{\phi}k^{\phi}, \label{K22} \\
	K_{33}&=(1-q C^{0303})g_{tt}k^tk^t+(1-q C^{0303})r^2k^{\phi}k^{\phi},
\end{align}
where we have applied $g_{tt}=-e^0_te^0_t$, $g_{rr}=e^1_re^1_r$ and $g_{\phi\phi}=e^3_{\phi}e^3_{\phi}$. The condition of non-zero solutions is $K_{11}K_{22}K_{33}=0$, which gives the below three roots and the corresponding solution of $a_{\mu}$:
\begin{itemize}
	\item $K_{11}=0$, $a_{\mu}\propto l_{\mu}$, $\vec{r}$ direction polarization,
	\item $K_{22}=0$, $a_{\mu}\propto m_{\mu}$, $\vec{\theta}$ direction polarization,
	\item $K_{33}=0$, $a_{\mu}\propto n_{\mu}$, unphysical polarization,
\end{itemize}
where we also write down the corresponding direction of linear polarization, i.e. the direction of electric field according to $E_i=\partial_tA_i-\partial_iA_t$. $K_{33}=0$ corresponds to the unphysical polarization since $n_{\nu}=k^0e^3_{\nu}-k^3e^0_{\nu}$ contains the non-transverse polarization component $e^3_{\nu}$.

As a result, one could find that photons with different linear polarizations sense different two dimensional ``effective metrics" on the circular orbits. For example, the photons with $\vec{r}$ polarization for $K_{11}=0$ corresponds to $g'_{tt}=(1-q C^{0101})g_{tt}$ and $g'_{\phi\phi}=(1+q C^{1313})r^2$. In order to be consistent with the motion in a planar circular orbits, the expansion coefficients $C^{abcd}$ should only depend on the coordinate $r$, which makes the effective metric component $g'_{\phi\phi}$ could in principle be set to $r^2$ by a redefinition of $r$, i.e.
\begin{align}
	ds^2=g_{tt}^s(u)dt^2+r^2d\phi^2,
	\label{am}
\end{align}
where the superscript $s$ represents different linear polarized photons. This kind of redefinition will not influence the observables according to the below Eq. (\ref{observable}). Eq. (\ref{am}) means photons with different polarizations will feel a different $g_{tt}$ component, i.e. the gravitational potential, which could therefore behave as the violation of WEP. As a result, one could test whether the new coupling term in the Lagrangian (\ref{La}) exists by checking whether photons with different linear polarizations sense different gravitational potentials.

\section{Black hole photon ring as a new probe of the EEP}
\label{as probe}

In the above discussions, we have pointed out for the photons in circular orbits, different linear polarized photons sense different gravitational potentials. In order to detect this novel phenomenon, one need to seek an observable corresponding to the gravitational potential. For the planar motion, let us consider the below $2+1$ planar metric
\begin{align}
	ds^2=g^s_{tt}(u)dt^2+g^s_{rr}(u)dr^2+r^2d\phi^2,
	\label{2dm}
\end{align}
where we have added the $rr$ component $g^s_{rr}(u)$ that satisfies $g^s_{rr}(0)=1$ comparing with (\ref{am}) since it is the direct approach to yield a circular orbit. Then the Lagrangian for the motion of photons is
\begin{align}
	\mathcal{L}=\frac{1}{2}g^s_{\mu\nu}(x)\dot{x}^{\mu}\dot{x}^{\nu},
	\label{L}
\end{align}
where $x$ is the spacetime coordinate and dot represents the derivative with respect to the affine parameter $\lambda$. There are three conservation quantities corresponding to the absence of $\lambda$, $t$ and $\phi$ in the Lagrangian respectively, i.e.
\begin{align}
	E&\equiv g_{tt}^s(r)\dot{t}, \\
	L&\equiv r^2\dot{\phi}, \\
	\zeta &\equiv g_{tt}^s(r)\dot{t}^2+g_{rr}^s(r)\dot{r}^2+r^2\dot{\phi}^2.
\end{align}
For the massless particle, $\zeta=0$. According to these three conservation quantities, we could obtain the equation of the trajectory $u(\phi)$
\begin{align}
	g^s_{tt}(u)g^s_{rr}(u)\left(\frac{du}{d\phi}\right)^2+u^2g^s_{tt}(u)=-\frac{E^2}{L^2}M^2.
	\label{eom}
\end{align}
For the conditions of bound orbits $du/d\phi=0$ and $d^2u/d\phi^2=0$, Eq. (\ref{eom}) could be simplified as the below two equations of $u$,
\begin{align}
	u^2g^s_{tt}(u)=-X^{-2}, \label{c1} \\
	2g^s_{tt}(u)+ug^s_{tt}{'}(u)=0, \label{c2}
\end{align}
where the prime represents the derivative with respect to $u$ and we have defined $X=L/(ME)$. As for the necessity for the existence of photon bound orbits, we refer readers to \cite{Claudel:2000yi} for more details.

The physical meaning of $X$ is the radius $d$ of the photon ring on the image plane divided by the mass $M$, whose expression is \cite{Chandrasekhar:1985kt}
\begin{align}
	\frac{d}{M}=\frac{1}{M}\lim_{r\rightarrow\infty}r^2\frac{d\phi}{dr}=X=\left[-u^2g^s_{tt}(u)\right]^{-\frac{1}{2}},
	\label{observable}
\end{align}
where we have used Eq. (\ref{eom}) and $g^s_{rr}(0)=-g^s_{tt}(0)=1$. Eq. (\ref{c2}) gives rise to the radius of the bound circular orbits and this radius will become the observed size of the photon ring by Eq. (\ref{c1}). Therefore, observations of photon ring would provide us with the direct information on the values of the gravitational potential $g^s_{tt}$ at the radius of the circular bound orbits and thus could let us know if the EEP violation coupling (\ref{La}) exists. Note that the $g^s_{rr}$ component will not affect the test due to its absence in Eq. (\ref{observable}).

\section{Method and Results}

In this section we will discuss the method to use black hole photon ring to constrain the EEP violation. According to the asymptotically flat condition $g^s_{tt}(0)=-1$, the metric component $g^s_{tt}(u)$ could be generally written as
\begin{align}
	g^s_{tt}(u)=-1+2u(1+\beta^s(u)),
	\label{gstt}
\end{align}
where $\beta^s(u)$ is the correction function for the Schwarzschild black hole induced by violation of the EEP. It could be formally expressed as
\begin{align}
	\beta^s(u)=\sum_{n}\beta_n^su^n,
	\label{beta}
\end{align}
where $n\geq-1$ so that the asymptotically flatness is satisfied and $n$ does not have to be an integer. The above expression effectively illustrates violation of the EEP could stem from any order of $M/r$, i.e. any strength of gravitational fields in principle.

 If $\beta^s(u)$ is small, we could decompose $u$ as the perturbed part $\delta u$ and the unperturbed part $u_0$, i.e. $u=u_0+\delta u$. And correspondingly, the size $X$ of the photon ring could be written as $X=X_0+\delta X$. The equation (\ref{c2}) up to the first order of $\beta^s$ and $\delta u$ gives
\begin{align}
	& 1-3u_0=0, \label{sp1} \\
	& -3\delta u=3u_0\beta^s(u_0)+u_0^2\beta^s{'}(u_0).
	\label{sp2}
\end{align}
So according to Eq. (\ref{c1}), we could obtain
\begin{align}
	\delta X=3\sqrt{3}\beta^s\left(u_0\right)
	\label{sdeltaX}
\end{align}
and $X_0=3\sqrt{3}$, where $u_0=1/3$ represents the radius of the unstable circular orbits of photons in the standard Schwarzschild solution. We can see that in the linear approximation, the deviation of $X$ has nothing to do with the derivative of $\beta^s(u)$ {and also the Eq. (\ref{sp2})}.

\begin{figure}
\includegraphics[width=.43\textwidth]{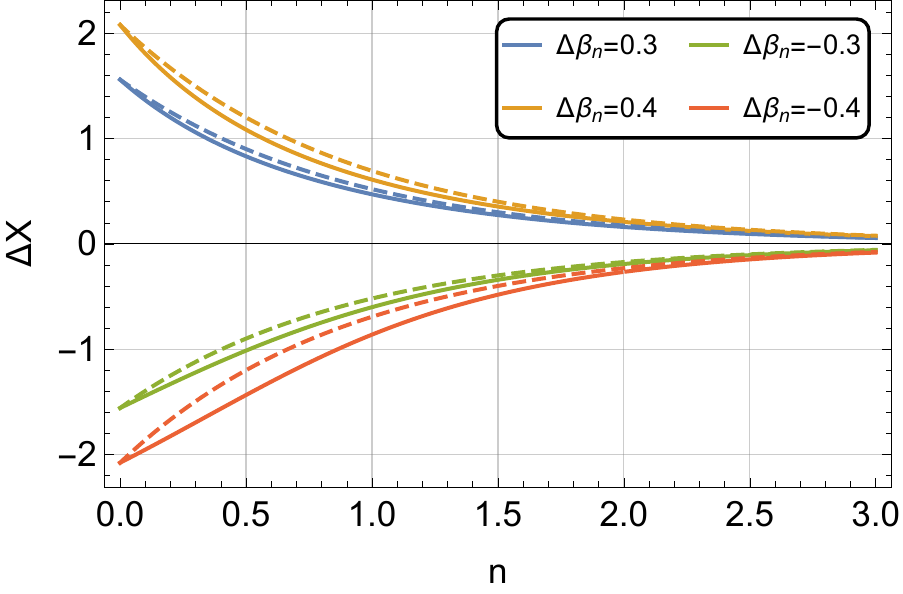}
\caption{Plot showing the difference in the size of the photon ring $\Delta X$ given by polarized $l$ and $m$ photons as $n$ changes for various $\Delta\beta_n$. The dashed lines correspond to the approximated expression (\ref{deltax}) and sold lines are results from rigidly solving Eq. (\ref{c1}) and Eq. (\ref{c2}).}
\label{deltaX}
\end{figure}

By observing the deviation of the photon ring's size from that of the standard general relativity $\delta X$, one could directly obtain the constraint on the magnitude of $\beta^s(u_0)$ according to Eq. (\ref{sdeltaX}). However, in order to give a decisive criterion whether the EEP is violated, we need at least two groups of photons with different linear polarizations. The observable is the difference of the photon ring's size presented by these two kinds of photons. After using $l$ and $m$ to label the corresponding $\vec{r}$ and $\vec{\theta}$ polarization, this difference according to Eq. (\ref{sdeltaX}) is given by
\begin{align}
	\Delta X=\delta X_l-\delta X_m&=3\sqrt{3}\left(\beta^l\left(u_0\right)-\beta^m\left(u_0\right)\right) \nonumber \\
	&\equiv3\sqrt{3}\Delta\beta\left(u_0\right),
	\label{deltax}
\end{align}
where we can see that in the linear approximation, $\Delta X$ only has dependence on the difference between $\beta^l$ and $\beta^m$ while the specific values of $\beta^l$ or $\beta^m$ does not have influence. By observing the magnitude of $\Delta X$, one could obtain the constraints on $\Delta\beta(u_0)$. Now we choose $\beta^s(u)=\beta^s_nu^n$, which gives $\Delta X=3^{\frac{3}{2}-n}\Delta\beta_n$ by Eq. (\ref{deltax}). In Fig. \ref{deltaX} we plot $\Delta X$ from $n=0$ to $n=3$ for different values of $\Delta\beta_n$ in dashed lines, where we can find that besides the proportional relation (\ref{deltax}), the effects of $\Delta\beta_n$ is suppressed as $n$ grows. This is caused by the suppression of $\beta^s(u_0)$ in Eq. (\ref{gstt}) for large values of $n$.

The precise results could be obtained by rigidly solving Eq. (\ref{c1}) and Eq. (\ref{c2}) for different values of $\beta^l_n$ and $\beta^m_n$, which are also shown in Fig. \ref{deltaX} through the solid line. We can find that the approximated expression (\ref{deltax}) could overestimate the values of $\Delta X$ and this overestimation is suppressed by large and small values of $n$. The reason of this suppression is that a larger $n$ tends to give a smaller value of $\beta^s(u_0)$ and $\beta^s(u_0)\approx\beta^s_n$ when $n$ approaches zero, which make the approximated equations (\ref{sp1}) and (\ref{sp2}) work better.

\begin{figure*}
\subfigure{
	\label{E1-1}
	\includegraphics[width=.31\textwidth]{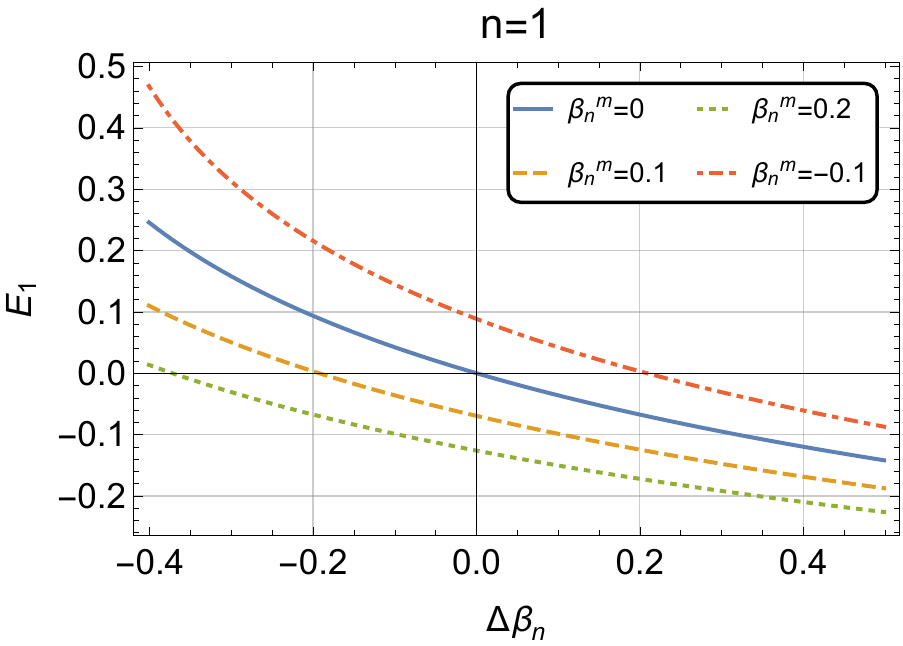}}
\subfigure{
	\label{E1-2}
	\includegraphics[width=.31\textwidth]{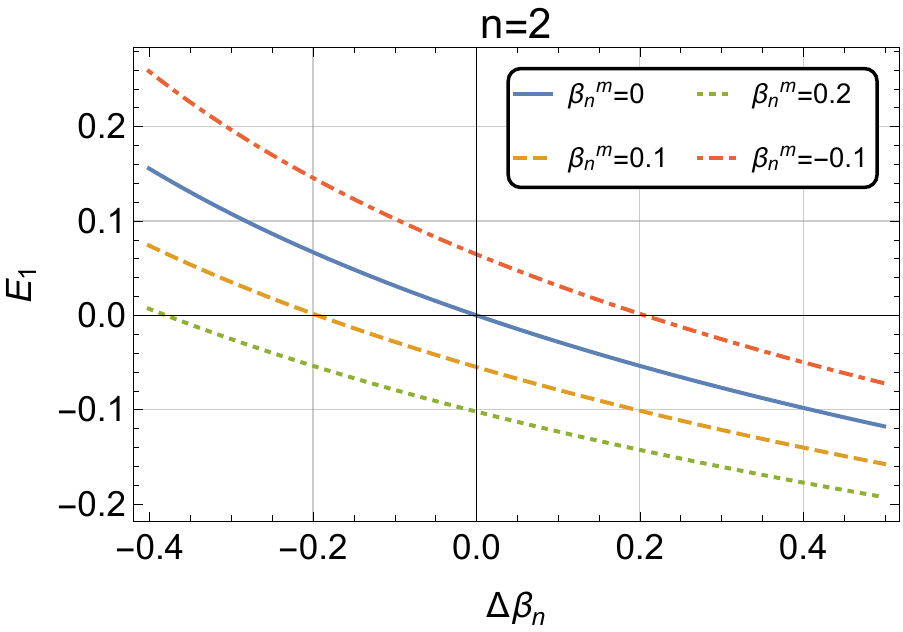}}
\subfigure{
	\label{E1-3}
	\includegraphics[width=.31\textwidth]{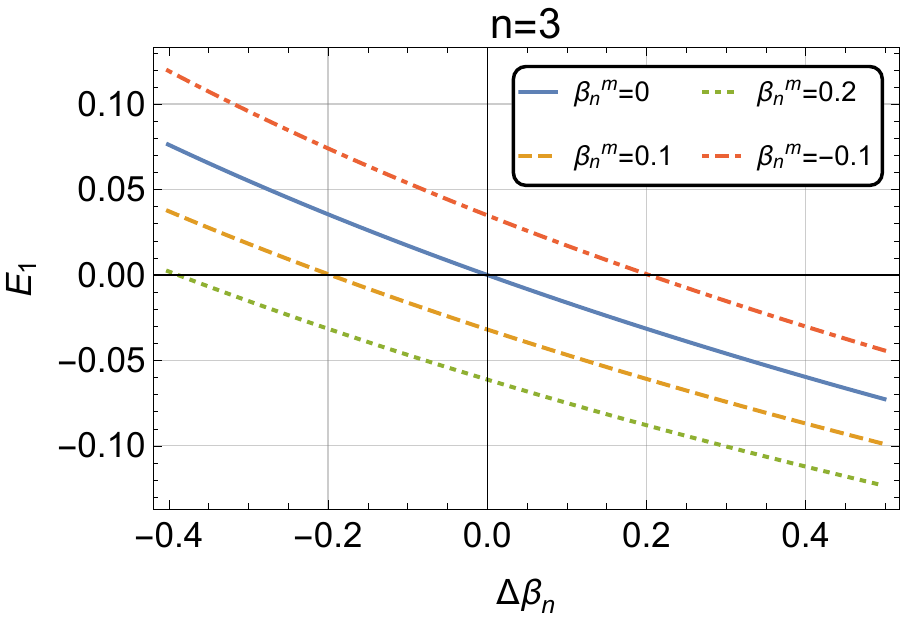}}
\caption{Plot showing the deviation ratio $E_1$ defined by Eq. (\ref{e1}) of the approximation (\ref{deltax}) for different values of $n$, $\Delta\beta_n$ and $\beta_n^m$.}
\label{delta1}
\end{figure*}

In order to better estimate the error of the proportional relation (\ref{deltax}), we define $E_1$ to measure the deviation ratio from the precise results, whose expression is
\begin{align}
	E_1\equiv\frac{\Delta X_N-\Delta X}{\Delta X},
	\label{e1}
\end{align}
where $\Delta X_N$ is the precise result from rigidly solving Eq. (\ref{c1}) and Eq. (\ref{c2}). In Fig. \ref{delta1} we show $E_1$ for different parameters so that one could intuitively obtain the conditions for the approximation (\ref{deltax}). Note that different from the approximated result, the specific value of $\beta^l$ or $\beta^m$ will matter in precise results. In this figure, besides the same information revealed by Fig. \ref{deltaX}, we could find that the positive and negative of $\beta_n^m$ have opposite effects on $E_1$ and as the absolute value of $\beta_n^m$ and $\Delta\beta_n$ grows, the difference between precise results and approximated results is enlarged, which is caused by the fact that the approximation (\ref{deltax}) only works well for the small EEP violation function $\beta^s(u)$.

Finally, we emphasize that although it is difficulty to obtain a general solution of photon ring's size $X$ as a function of any order of the EEP violation parameters $\beta^s_n$ and $n$ in Eq. (\ref{beta}), one could also obtain a general expression of photon ring's size $X$ through a matching method by not specifying a form of the EEP violation function $\beta^s(u)$. This matching method is based on the assumption that the deviation of the radius of circular orbits from that of general relativity is relatively small. Specifically, one could firstly define a function $B(u)=b_pu+b_0+b_nu^{-1}$. After replacing $\beta^s(u)$ with $B(u)$, Eq. (\ref{c1}) and Eq. (\ref{c2}) will have the below analytical solution for $X^{-2}$
\begin{align}
	X^{-2}(b_p,b_0,b_n)=-\frac{2(-1+2b_n)^3\left(1+\sqrt{9+\frac{16(1-2b_b)b_p}{(1+b_0)^2}}\right)}{(1+b_0)^2\left(3+\sqrt{9+\frac{16(1-2b_b)b_p}{(1+b_0)^2}}\right)^3}.
	\label{xm2}
\end{align}
For any target function $\beta^s(u)$, we could match the zero, first, third order derivative of $B(u)$ to those of $\beta^s(u)$ at $u=1/3$, which leads to the below expressions for the coefficients of $B(u)$
\begin{align}
	b_p&=\beta^s{'}\left(\frac13\right)+\frac16\beta^s{''}\left(\frac13\right), \\
	b_0&=\beta^s\left(\frac{1}{3}\right)-\frac13\beta^s{'}\left(\frac{1}{3}\right)-\frac{1}{9}\beta^s{''}\left(\frac{1}{3}\right), \\
	b_n&=\frac{1}{54}\beta^s{''}\left(\frac{1}{3}\right).
\end{align}
Then by substituting the above coefficients into Eq. (\ref{xm2}), one could obtain an approximated expression of the photon ring's size $X$ for the target function $\beta(u)$. For example, $\beta^s(u)=\beta_0+\beta_1u+\beta_2u^2$ corresponds to $X^{-2}(\beta_1+\beta_2,\beta_0-1/3\beta_2,1/27\beta_2)$. And if we choose $\beta_0=1/2$, $\beta_1=-3/2$ and $\beta_2=1/2$, the fitting method gives $X=5.375$ while the rigid solution gives $X=5.379$ with an error only $0.075\%$. And for the above selected target EEP violation function $\beta^s(u)=\beta_nu^n$, one could verify that the maximum deviation ratio of the above fitting method from the rigid solutions is around $0.1\%$ for the case of a $50\%$ change in photon ring's size.

\section{Examples}

In this section, we present several examples to illustrate how to get the EEP violation parameters $\beta^s_n$ from a specific model.

\subsection{Vector field}

Now we consider a vector field $V^{\mu}$ being coupled to electromagnetic tensor $F_{\mu\nu}$. Given that the index symmetry of $Q^{\mu\nu\rho\sigma}$, the corresponding expression of $Q^{\mu\nu\rho\sigma}$ is
\begin{align}
	Q^{\mu\nu\rho\sigma}=\frac{1}{2}g^{\nu[\sigma}V^{\rho]}V^{\mu}+\frac{1}{2}g^{\mu[\rho}V^{\sigma]}V^{\nu}.
\end{align}
As an example, we choose the Schwarzschild spacetime. If the vector field $V^{\mu}$ has the same symmetry as the spacetime, i.e. the static spherical symmetry, one of the allowed forms of $V^{\mu}$ is $(0,v(r),0,0)$. Therefore, the nonzero expansion coefficients according to Eq. (\ref{expansion}) are
\begin{align}
	C^{1212}=C^{1313}=-C^{0101}=f(r),
	\label{ve}
\end{align}
where the definition of $f(r)$ is $f(r)=1/4V^1V^1$ and $V^1=V^{\mu}e^1_{\mu}$. Now let us choose a specific expression $f(r)=1/r^2$. For the planar circular orbits $k^{\theta}=0$, $k^r=0$ and $\theta=\pi/2$, the dispersion relation (\ref{K11}) corresponding to $\vec{r}$ polarized photons gives the below gravitational potential
\begin{align}
	g^l_{tt}=-1+\frac{2M}{r}-q\frac{1}{r^2}+3q\frac{M}{r^3},
\end{align}
where we have eliminated the $g_{\phi\phi}$ component by the redefinition of $r$ and only retained the first order terms with respect to $q$. For $\vec{\theta}$ polarized photons, the gravitational potential is not modified, i.e. $\beta^m_n=0$. Then the expansion coefficients of Eq. (\ref{beta}) for $\vec{r}$ polarized photons are
\begin{align}
	\beta^l_1=-\frac{q}{2M^2},\ \beta^l_2=\frac{3q}{2M^2}.
\end{align}
The existence of the planar circular orbits could be verified by the matrix (\ref{system}). The condition of nonzero solutions is reduced to $K_{11}K_{22}K_{33}=0$ for the expansion coefficients (\ref{ve}) and the expressions of $K_{11}$, $K_{22}$ and $K_{33}$ are
\begin{align}
	K_{11}&=(1+q f)(g_{tt}k^tk^t+g_{rr}k^rk^r+g_{\theta\theta}k^\theta k^\theta \nonumber \\
	&+g_{\phi\phi}k^\phi k^\phi), \\
	K_{22}&=K_{33}=g_{tt}k^tk^t+(1+q f)g_{rr}k^rk^r+g_{\theta\theta}k^\theta k^\theta \nonumber \\
	&+g_{\phi\phi}k^{\phi\phi}.
\end{align}
Therefore, for each of the situations $K_{11}=0$, $K_{22}=0$ or $K_{33}=0$, photons will have a dispersion relation described by a modified Schwarzschild metric, which still has static spherical symmetry and thus leads to the existence of the planar circular orbits. The planar motion is described by the metric (\ref{2dm}).

\subsection{Tensor field}

We choose $Q^{\mu\nu\sigma\rho}=R^{\mu\nu\sigma\rho}$, which describes the correction of the virtual electron loops on the propagation of photons \cite{Daniels:1995yw,Lafrance:1994in}. For the Schwarzschild spacetime, the nonzero expansion coefficients of $R^{\mu\nu\sigma\rho}$ are
\begin{align}
	C^{0101}=-\frac{2M}{r^3}, \ C^{0202}=\frac{M}{r^3}, \ C^{0303}=\frac{M}{r^3}, \nonumber \\ C^{1212}=-\frac{M}{r^3}, \ C^{1313}=-\frac{M}{r^3}, \ C^{2323}=\frac{2M}{r^3}.
\end{align}
Under the condition of planar circular orbits $k^{\theta}=0$, $k^r=0$ and $\theta=\pi/2$, the dispersion relations (\ref{K11}) and (\ref{K22}) corresponding to the $\vec{r}$ and $\vec{\theta}$ polarized photons respectively lead to
\begin{align}
	g^l_{tt}&=-1+\frac{2M}{r}-2q\frac{M}{r^3}+3q\frac{M^2}{r^4}, \\
	g^m_{tt}&=-1+\frac{2M}{r}+q\frac{M}{r^3},
\end{align}
where we have eliminated the $g_{\phi\phi}$ components by the redefinition of $r$ and only retained the first order terms with respect to $q$. Therefore, the corresponding expansion coefficients of Eq. (\ref{beta}) are
\begin{align}
	&\beta^l_2=-\frac{q}{M^2},\ \beta^l_3=\frac{3q}{2M^2}, \\
	&\beta^m_2=\frac{q}{2M^2}.
\end{align}
As for the existence of the planar circular orbits, one could directly calculate the determinant of the matrix (\ref{system}), the condition of nonzero solutions gives rise to $K_{11}K'_{22}K'_{33}=0$ and the expressions of $K_{11}$, $K'_{22}$ and $K'_{33}$ are
\begin{align}
	K_{11}&=\left(1+q\frac{2M}{r^3}\right)\left(g_{tt}k^tk^t+g_{rr}k^rk^r\right) \nonumber \\
	&+\left(1-q\frac{M}{r^3}\right)\left(g_{\theta\theta}k^{\theta}k^{\theta}+g_{\phi\phi}k^{\phi}k^{\phi}\right), \\
	K'_{22}&=\left(1-q\frac{M}{r^3}\right)\left(g_{tt}k^tk^t+g_{rr}k^rk^r\right) \nonumber \\
	&+\left(1+q\frac{2M}{r^3}\right)\left(g_{\theta\theta}k^{\theta}k^{\theta}+g_{\phi\phi}k^{\phi}k^{\phi}\right), \\
	K'_{33}&=\left(1-q\frac{M}{r^3}\right)(g_{tt}k^tk^t+g_{rr}k^rk^r \nonumber \\
	&+g_{\theta\theta}k^{\theta}k^{\theta}+g_{\phi\phi}k^{\phi}k^{\phi}).
\end{align}
Similar to the vector situation, for each of the conditions $K_{11}=0$, $K'_{22}=0$ or $K'_{33}=0$, the dispersion relation is described by a static metric with spherical symmetry, which leads to the existence of the planar circular orbits and the planar motion is described by the metric (\ref{2dm}).

\subsection{Scalar field}

For the scalar coupling such as $Q^{\mu\nu\rho\sigma}=2f(\phi)g^{\rho[\mu}g^{\nu]\sigma}$\cite{Sandvik:2001rv,Bekenstein:1982eu} as discussed earlier, one could directly verify that the standard dispersion relation $k^{\mu}k_{\mu}=0$ is not modified according to Eq. (\ref{dr}). If $\phi$ is the axion field, when $Q^{\mu\nu\rho\sigma}=-\phi\xi^{\mu\nu\rho\sigma}$ and $\xi^{\mu\nu\rho\sigma}$ is the Levi-Civita symbol \cite{Schwarz:2020jjh}, the standard dispersion relation $k^{\mu}k_{\mu}=0$ is still preserved. In order to directly test this kind of coupling, one could look for the higher order geometric optics approximation, which leads to the rotation of polarization vectors along the path of photons and could be tested by the precise measurements of polarizations in principle \cite{Chen:2019fsq}. For other forms of tensor coupling, we refer readers to \cite{Chen:2020qyp,Huang:2016qnl} for more details.

\section{The influence of the rotation of black hole}
\label{b}

In the above, we have discussed the EEP violation for the spacetime of static spherically symmetric black hole. The only metric component associated with the photon ring observations is $g^s_{tt}$, which thus could let us exclude the influence of other metric components and only focus on the EEP violation manifested as different gravitational potentials. However, when the black hole has rotation, the situation will become complicated. Due to the diminution of spacetime symmetries, one could expect that apart from $g_{tt}$, other metric components such as $g_{t\phi}(r,\theta)$ will become relevance. This fact makes the above model independent discussions based on the assumption of planar circular orbits become difficult to carry on. Furthermore, for rotating black holes, the spin parameter and the inclination angle of rotation axis also need to be determined in order to give a precise prediction of photon ring. All of these factors add more parameters to the process of constraining the EEP violation and thus increase the complexity. Fortunately, the detailed numerical study based on the Kerr black hole shows that new parameters introduced by the rotation, which include the spin parameter and the inclination angle of rotation axis, mainly affect the horizontal displacement and the outline's asymmetry of photon ring. While the overall size of ring could hardly be changed by rotation effects \cite{Chan:2013tsa,Johannsen:2010ru}.

Therefore, in this section we focus on the overall size of the black hole photon ring. What we are interested in is how much the rotation of black holes affects the results of constraining the EEP violation. Our strategy is to characterize the black hole rotation by directly generalizing the effective metric (\ref{2dm}). Then we derive the relation between the observable and the EEP violation parameters similar to Eq. (\ref{deltax}). And based on this we study the influence of various parameters related to rotation on constraint results.

\subsection{The model}

According to the effective metric (\ref{2dm}) and the expression (\ref{gstt}), we start from the following Schwarzschild-like metric
\begin{align}
	ds^2&=-\left[1-2u\left(1+\beta^s(u)\right)\right]dt^2+\left[1-2u\left(1+\beta^s(u)\right)\right]^{-1}dr^2 \nonumber \\&+r^2\left(d\theta^2+\sin^2\theta\phi^2\right).
\end{align}
A solution of rotating black hole could be obtained by applying the Newman-Janis algorithm \cite{Newman:1965tw}. The above metric could correspond to the below rotating one in the Boyer-Lindquist coordinates $(t,r,\theta,\phi)$ \cite{Bambi:2013ufa}
\begin{align}
	ds^2&=-\left(1-\frac{2u\left(1+\beta^s(u)\right)}{1+A^2u^2\cos^2\theta}\right)dt^2 \nonumber  \\
	&+\frac{1+A^2u^2\cos^2\theta}{1-2u\left(1+\beta^s(u)\right)+A^2u^2}dr^2 \nonumber \\
	&+\frac{M^2}{u^2}\left(1+A^2u^2\cos^2\theta\right)d\theta^2 \nonumber \\
	&+\frac{M^2}{u^2}\left(1+A^2u^2+\frac{2A^2\left(1+\beta^s(u)\right)u^3\sin^2\theta}{1+A^2u^2\cos^2\theta}\right)\sin^2\theta d\phi^2 \nonumber \\
	&-M\frac{4\left(1+\beta^s(u)\right)Au\sin^2\theta}{1+A^2u^2\cos^2\theta}d\phi dt,
	 \label{metric}
\end{align}
where $A=a/M$ and $a$ is the angular momentum per unit mass of black hole, i.e. $a=J/M$. When $\beta^s$ vanishes, this will lead to the standard Kerr black hole.

The equation of motion could be obtained by the Hamilton-Jacobi equation:
\begin{align}
	H+\frac{\partial S}{\partial \lambda}=0,
	\label{hj}
\end{align}
where $S$ is the Hamilton principal function $S(\lambda,x^{\mu})$ and $\lambda$ is the affine parameter. The Hamiltonian $H$ is
\begin{align}
 H = -\frac{1}{2} g^{\mu\nu} P_{\mu} P_{\nu}
\label{Hamiltonian}
\end{align}
and $P_{\mu}$ in the Hamilton-Jacobi formalism is
\begin{align}
	P_{\mu}=\frac{\partial S}{\partial x^{\mu}}.
\end{align}
The solution of Eq. (\ref{hj}) with the metric (\ref{metric}) is separable and we refer readers to \cite{Chandrasekhar:1985kt,Abdujabbarov:2016hnw} for more details. The conditions of bound orbits $\dot{r}=0$ and $\ddot{r}=0$ with the solution give rise to
\begin{align}
	&(u^{-2}+A^2-Ax)^{2} \nonumber \\
	&-[u^{-2}-2u^{-1}(1+\beta^s(u))+A^2][y^2+(x-A)^2]=0, \label{k1} \\
	&2u^{-1}(u^{-2}+A^2-Ax) \nonumber \\
	&-(u^{-1}+u\beta^s{'}(u)-1-\beta^s(u))(y^2+(x-A)^2)=0, \label{k2}
\end{align}
where the definitions of $x$ and $y^2$ are
\begin{align}
	x&=\frac{L_z}{EM}, \\
	y^2&=\frac{\mathcal{K}}{E^2M^2}.
\end{align}
$E$ and $L_z$ are the integral constants corresponding to the absences of $t$ and $\phi$ in the Hamiltonian (\ref{Hamiltonian}) respectively. $\mathcal{K}$ is Carter constant which is introduced by the separability of the Hamilton-Jacobi equation (\ref{hj}). From Eq. (\ref{k1}) and Eq. (\ref{k2}), one could obtain the solutions of $x$ and $y$ with respect to $u$, i.e. $x(u)$ and $y(u)$.

For the rotating black holes, we need two coordinates $X$ and $Y$ on the image plane to describe the appearance of the black hole photon ring. These two parameters are equivalent to the initial conditions of light rays and in the unit of mass are given by \cite{Chandrasekhar:1985kt}
\begin{align}
	&X=-\frac{1}{\sin\theta_0}x, \label{X} \\
	&Y=\pm\left[y^2+(A-x)^2-\left(A\sin\theta_0-\frac{x}{\sin\theta_0}\right)^2\right]^{\frac12}, \label{Y}
\end{align}
where $\theta_0$ is the inclination angle between the rotation axis of the black hole and the line of sight of the distant observer. Substituting the solutions $x(u)$ and $y(u)$ of Eq. (\ref{k1}) and Eq. (\ref{k2}) into the expressions (\ref{X}) and (\ref{Y}), one could obtain the photon ring outline as the functions of $u$, i.e. $X(u)$ and $Y(u)$ in the domain having solutions.

\subsection{Method and results}

Similar to the discussion of static spherically symmetric black holes, we decompose $u$ as $u=u_0+\delta u$ in the condition that $\beta^s(u)$ is small. Then the first order equations with respect to $\delta u$ and $\beta$ from Eq. (\ref{k1}) and Eq. (\ref{k2}) are
\begin{align}
	&u_0\left[2u_0(A-x_0)+x_0\right]\delta x+u_0(1-2u_0+A^2u_0^2)y_0\delta y \nonumber \\
	&+\left\{2\left[u_0^{-2}+A(A-x_0)\right]+(u_0-1)\left[(A-x_0)^2+y_0^2\right]\right\}\delta u \nonumber \\
	&-u_0^2\left[(A-x_0)^2+y_0^2\right]\beta^s(u_0)=0, \label{u1}\\
	&2\left[(1-u_0^{-1})x_0-A\right]\delta x+2(1-u_0^{-1})y_0\delta y \nonumber \\
	&-u_0^{-4}\left[6+u_0^2(A^2-x_0^2-y_0^2)\right]\delta u+\left[(A-x_0)^2+y_0^2\right]\beta^s(u_0) \nonumber \\
	&-u_0\left[(A-x_0)^2+y_0^2\right]\beta^s{'}(u_0)=0, \label{u2}
\end{align}
where $\delta x$ and $\delta y$ represent the perturbed part of $x$ and $y$. $x_0$, $y_0$ and $u_0$ satisfy the zeroth order equations:
\begin{align}
	&(u_0^{-2}+A^2-Ax_0)^{2} \nonumber \\
	&-[u_0^{-2}-2u_0^{-1}+A^2][y_0^2+(x_0-A)^2]=0, \label{01} \\
	&2u_0^{-1}(u_0^{-2}+A^2-Ax_0) \nonumber \\
	&-(u_0^{-1}-1)[y_0^2+(x_0-A)^2]=0. \label{02}
\end{align}
Solving Eq. (\ref{01}) and Eq. (\ref{02}) to obtain $x_0(u_0)$ and $y_0(u_0)$ and substituting them into Eq. (\ref{u1}) and Eq. (\ref{u2}), we could obtain $\delta x$ and $\delta y$ as the functions of $u_0$ and $\delta u$, i.e. $\delta x(u_0, \delta u)$ and $\delta y(u_0, \delta u)$.

\begin{figure}
\includegraphics[width=.43\textwidth]{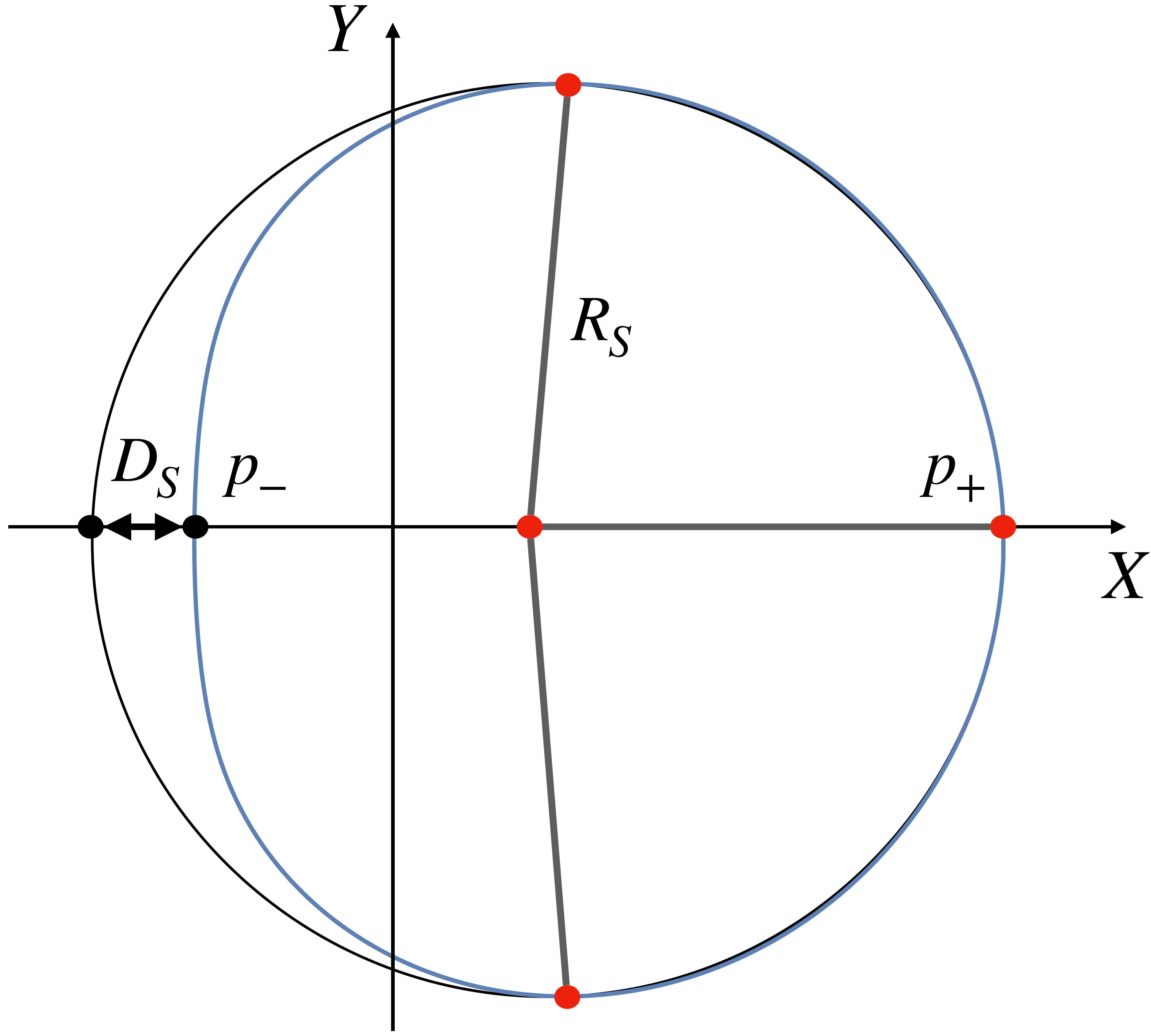}
\caption{Plot illustrating the definition and the geometrical meaning of the observables $R_S$ and $D_S$, where the blue line outlines the contour of the photon ring.}
\label{example}
\end{figure}

Because of the symmetry described by the metric (\ref{metric}), the outline of the photon ring always has a symmetry axis on the image plane which is implied by the sign of the expression (\ref{Y}). Therefore, one important feature of the photon ring relevant to the observations is the two intersections of the ring contour with the symmetry axis, which is shown as $p_+$ and $p_-$ in Fig. (\ref{example}) with $\pm$ be the positive and negative values of $X$. These two points are determined by the solutions of $Y=0$. Specifically, according to the expression (\ref{Y}), $Y$ could be written as
\begin{align}
	Y(x,y)=Y_0(x_0,y_0)+\delta Y(\delta x,\delta y).
\end{align}
Substituting $\delta x(u_0, \delta u)$ and $\delta y(u_0, \delta u)$ into $\delta Y$ and solving $\delta Y=0$, we could obtain $\delta u$ at the points of $p_+$ and $p_-$ respectively, i.e. $\delta u_{\pm}(u_{0\pm})$ where $u_{0\pm}$ are the solutions of the zeroth order equation $Y_0=0$.

According to Eq. (\ref{X}) and the expressions of $\delta u_{\pm}(u_{0\pm})$, the perturbed intercepts $\delta X_{\pm}$ between the contour of the photon ring and the symmetry axis on the image plane are
\begin{align}
	\delta X_{\pm}(u_{0\pm})&=-\frac{1}{\sin\theta_0}\delta x_{\pm}(u_{0\pm},\delta u_{\pm}) \nonumber \\
	&=f(A,\theta_0,u_{0\pm})\beta^s(u_{0\pm}),
	\label{pm}
\end{align}
where we have defined
\begin{align}
	&f(A,\theta_0,u_{0\pm})\equiv \nonumber \\
	&\frac{-4A\sin\theta_0 u_{0\pm}}{\left(1-u_{0\pm})(1-3u_{0\pm}+A^2\cos^2\theta_0 u_{0\pm}^2+A^2\cos^2\theta_0 u_{0\pm}^3\right)}.
\end{align}
Similar to the black hole without rotation, $\delta X_{\pm}$ is proportional to $\beta^s$ at the linear approximation. As for the unperturbed parts $X_{0\pm}$, they could be obtained by replacing $x$ in Eq. (\ref{X}) with the solution $x_0(u_0)$ of Eqs. (\ref{01}, \ref{02}) and let $u_0$ equal to $u_{0\pm}$.

Now let us consider the observables that characterize the photon ring contour. We use the method developed in \cite{Hioki:2009na} where two observables $R_S$ and $D_S$ are defined to quantify the size and the distortion in shape of the photon ring respectively. This method is based on the assumption that the outline of the black hole photon ring is nearly circular, which is true for the standard Kerr black hole and various of rotating black holes with separable geodesics including the model (\ref{metric}). According to the way in which we defined the coordinates $\alpha$ and $\beta$, the general shape and position of the photon ring on the image plane are shown in Fig. \ref{example} as the blue line. One can always draw a reference circle which is uniquely defined by three points: the top, bottom and the rightmost points as shown in Fig. \ref{example} with black line. The first observable $R_S$ is the radius of the reference circle which describes the apparent overall size of the photon ring. The second observable $D_S$ is defined by the apparent distance between leftmost point of the ring contour and that of the reference circle, which thus measures the degree of the ring contour deviating from a perfect circle.

We denote the coordinate of the center of the reference circle to $(C,0)$. When the appearance of the photon ring is changed by the EEP violation term $\beta^s(u)$, the center of the reference circle will also be changed along the axis of $\alpha$. Therefore, we use $\delta C$ to label the changed part and $C_0$ to label the original part. The expressions of $R_S$ and $D_S$ could be written as
\begin{align}
	R_S&=X_{+}-C=X_{0+}+\delta X_+-C_0-\delta C, \\
	D_S&=X_{+}-\vert X_{-}\vert-2C \nonumber \\
	&=X_{0+}+\delta X_+-\vert X_{0-}+\delta X_-\vert-2C_0-2\delta C.
\end{align}
In this paper, we only focus on the overall size $R_S$ since it is a much more obvious observable than the deformation $D_S$ and the above approximation works well. For two groups of photons $l$ and $m$ with different linear polarizations, we have
\begin{align}
	\Delta R_S&=R_S^l-R_S^m=\delta X_+^l-\delta X_+^m-\delta C^l+\delta C^m \nonumber \\
	&\approx\delta X_+^l-\delta X_+^m=f(A,\theta,u_{0+})\Delta\beta(u_{0+}),
	\label{deltar}
\end{align}
where $\Delta\beta(u_{0+})=\beta^l(u_{0+})-\beta^m(u_{0+})$ and we have ignored the difference of $\delta C$ between $l$ and $m$ since the detailed numerical study shows that small $\beta^s$ makes the top, bottom and the rightmost points in Fig. \ref{example} almost have the same magnitude of displacement, which makes the change of the circle center be sub-dominated. $\Delta R_S$ is proportional to $\Delta\beta$ and does not depend on the specific values of $\beta^l$ or $\beta^m$, which is similar to the situation without rotation. 

\begin{figure}
\includegraphics[width=.43\textwidth]{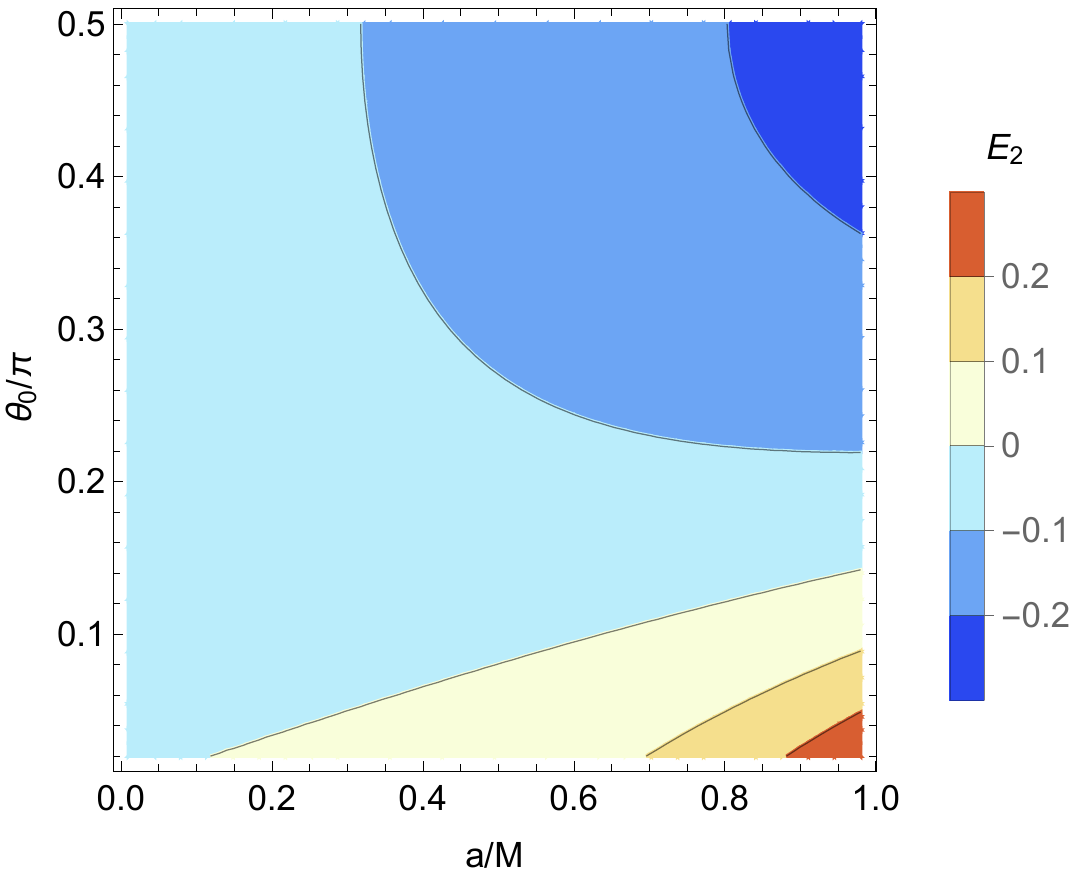}
\caption{The figure shows that the variation of $E_2$ defined by Eq. (\ref{e2}) as different values of the spin parameter $a$ and the inclination angle $\theta_0$ for $n=1$.}
\label{E2}
\end{figure}

\begin{figure}
\includegraphics[width=.43\textwidth]{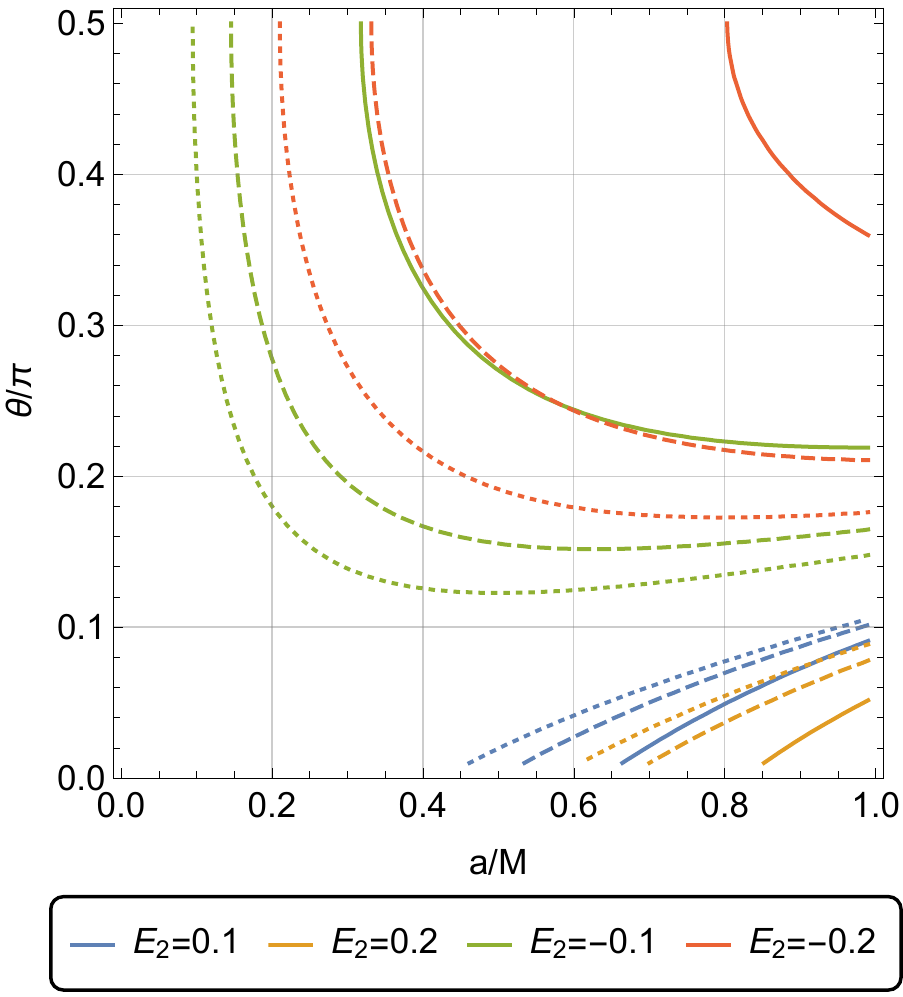}
\caption{Similar to Fig. \ref{E2}, this figure shows contour lines of different values of $E_2$. The solid, dashed, dot lines correspond to $n=1,2,3$ respectively. The colors represent values of $E_2$.}
\label{E2p}
\end{figure}

In order to characterize the influence of the black hole rotation, we define the fraction of the change in the overall size caused by the rotation as
\begin{align}
	E_2=\frac{\Delta R_S-\Delta X}{\Delta X}=\frac{f(A,\theta,u_{0+})u_{0+}^n-3\sqrt{3}u_0^n}{3\sqrt{3}u_0^n},
	\label{e2}
\end{align}
where $\Delta X$ is the result of no rotation situation, i.e. Eq. (\ref{deltax}) and we have chosen $\beta^s(u)=\beta^s_nu^n$. In Fig. \ref{E2}, we plot the value of $E_2$ as the variation of the spin parameter $a/M$ and the inclination angle $\theta_0$ for $n=1$. We could see that the largest deviation ratio from the no rotation black hole occurs in the largest $a$ and in the nearly face-on or edge-on view corresponding to $\theta_0=0$ and $\theta_0=\pi/2$ respectively. Then in Fig. \ref{E2p}, for different values of $n$, we plot the corresponding contour lines of different values of $E_2$. The solid, dashed, dot lines correspond to $n=1,2,3$ respectively. We can see that a larger $n$ tends to produce a large deviation ratio from the no rotation case. The reason is that for a given value of $\Delta\beta_n$, a larger $n$ means a smaller effect of the EEP violation which will be more comparable with the effects caused by the rotation.

\begin{figure*}
\subfigure[{$a/M=0.3$, $\theta_0=\pi/2$, $n=1$, $\beta_n^c=0$}]{
	\label{E3-1}
	\includegraphics[width=.44\textwidth]{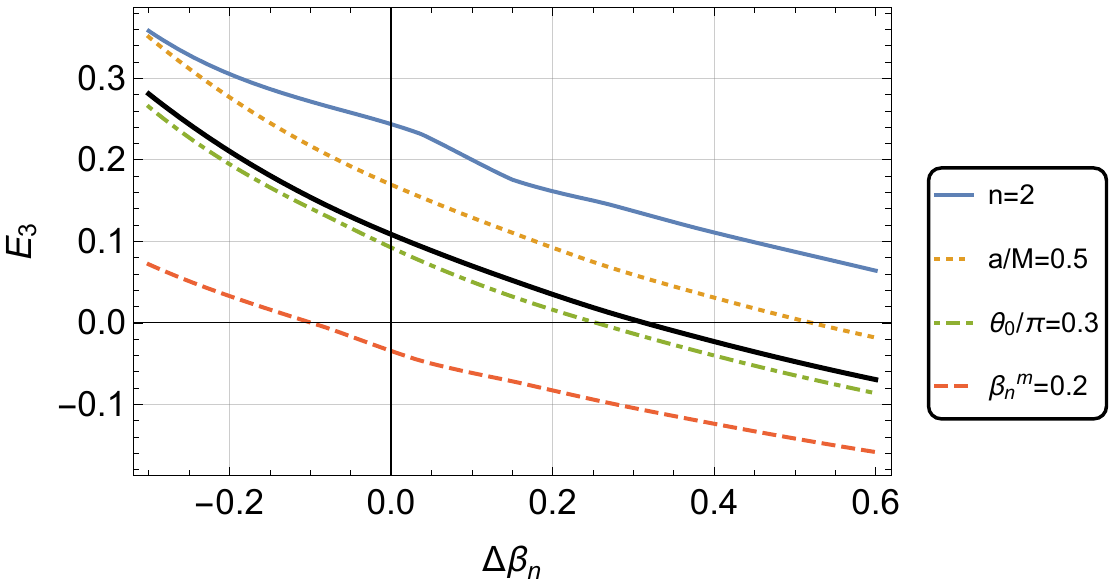}}
\subfigure[{$a/M=0.3$, $\theta_0=\pi/2$, $n=1$, $\Delta\beta_n=0.2$}]{
	\label{E3-5}
	\includegraphics[width=.43\textwidth]{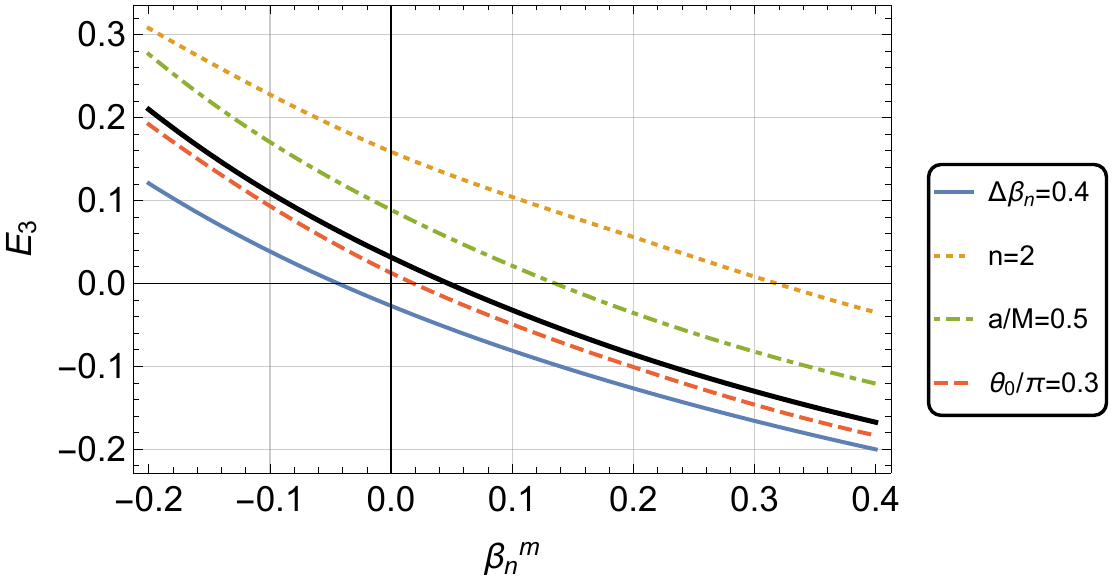}}	
\subfigure[{$a/M=0.3$, $\theta_0=\pi/2$, $\Delta\beta_n=0.2$, $\beta_n^c=0$}]{
	\label{E3-2}
	\includegraphics[width=.44\textwidth]{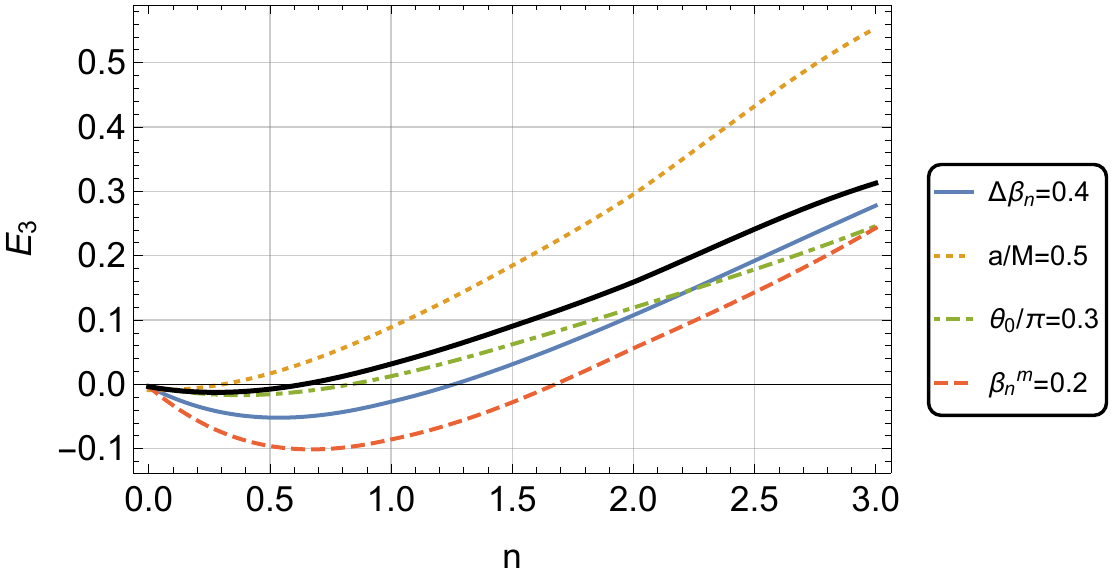}}
\subfigure[{$\theta_0=\pi/2$, $n=1$, $\Delta\beta_n=0.2$, $\beta_n^c=0$}]{
	\label{E3-3}
	\includegraphics[width=.44\textwidth]{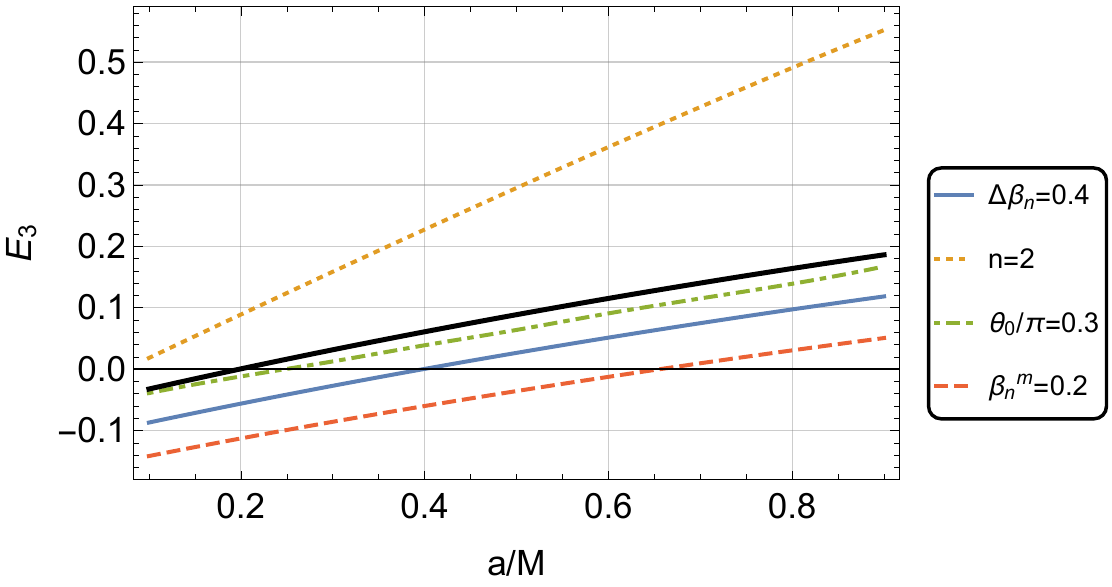}}
\subfigure[{$a/M=0.3$, $n=1$, $\Delta\beta_n=0.2$, $\beta_n^c=0$}]{
	\label{E3-4}
	\includegraphics[width=.44\textwidth]{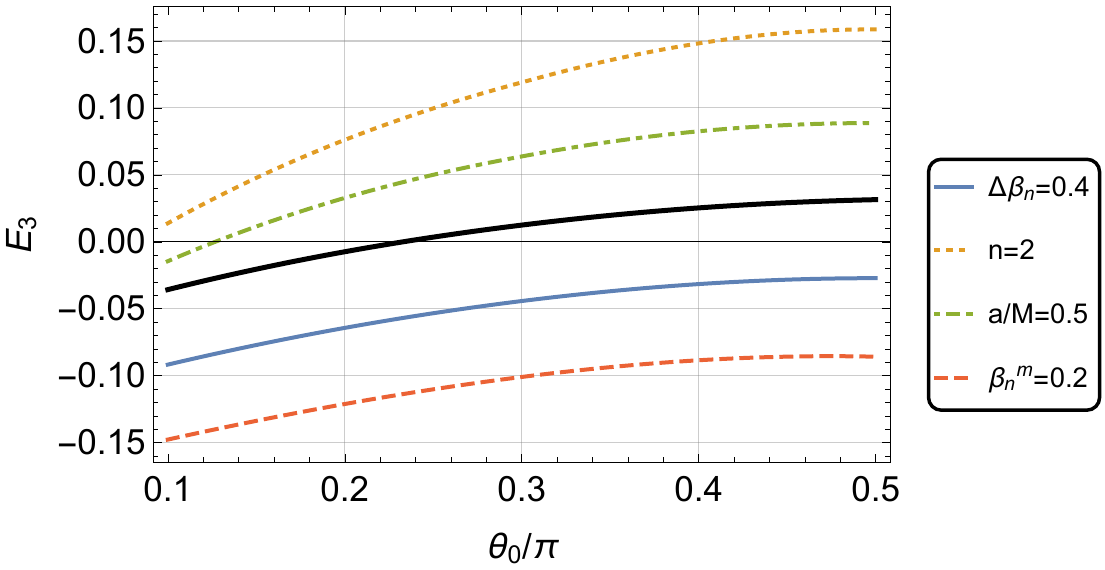}}
\caption{Plot showing the deviation ratio $E_3$ which is defined by Eq. (\ref{e3}) as the variation of all kinds of parameters. The parameters corresponding to the black curve are shown in below of each figure. The legends represent different selections of parameter that are changed comparing with the black curve.}
\label{E3}
\end{figure*}

Finally, we compare the fully numerical results with the approximated expression (\ref{deltar}) by defining
\begin{align}
	E_3=\frac{\Delta R_S^N-\Delta R_S}{\Delta R_S},
	\label{e3}
\end{align}
where $\Delta R_S^N=R^{Nl}_S-R^{Nm}_S$ denotes the numerical results, i.e. $R^{Nl}_S$ and $R^{Nm}_S$ are obtained from the contour of photon ring by numerically solving Eq. (\ref{k1}) and Eq. (\ref{k2}). The nonzero values of $E_3$ have two sources. The first one is the linear approximation (\ref{pm}) by assuming the EEP violation function $\beta(u)$ is small. The second one is the approximated expression (\ref{deltar}) of $\Delta R_S$ characterizing the overall size of photon ring. In Fig. \ref{E3}, we plot the deviation ratio $E_3$ as the variation of all kinds of parameters including the spin parameter $a$, the inclination angle $\theta_0$ and the parameters related to the EEP violation $\Delta\beta_n$, $n$ and $\beta_n^m$. In each figure of Fig. \ref{E3}, the parameters corresponding to the black curve are shown in below. Other colored curves are results of different selections of the parameter that are changed comparing with the black one and the changed parameters are shown in the legends. A significant feature of Fig. \ref{E3} is that the deviation ratio $E_3$ is approximately proportional to all parameters except for $n$ and the proportionality coefficients are approximately independent of any other parameters other than $n$. Therefore, in order to let readers better estimate the magnitude of the deviation ratio $E_3$, we use proportional expressions to describe the trend of the black lines in Fig. \ref{E3-1}, \ref{E3-5}, \ref{E3-3} and \ref{E3-4} respectively, i.e.
\begin{itemize}
	\item (a): $E_3=-0.39\Delta\beta_n+0.16$,
	\item (b): $E_3=-0.63\beta^m_n+0.08$,
	\item (d): $E_3=0.26(a/M)-0.06$,
	\item (e): $E_3=0.18(\theta_0/\pi)-0.06$.
\end{itemize}

From Fig. \ref{E3-1} and Fig. \ref{E3-5}, one could find that similar to the no rotation case (\ref{e1}), the absolute value of $E_3$ increases with that of $\Delta\beta_n$ and $\beta_n^m$. As for another EEP violation parameter $n$, from Fig. \ref{E3-2} one could see that $E_3$ tends to vanish as $n$ goes to zero. The reason for this is that the approximations (\ref{pm}) and (\ref{deltar}) work well for small $n$ ($n\lesssim 0.5$). While for large $n$ ($n\gtrsim 2$), the approximated expression (\ref{deltar}) works badly since the small change of photon ring size is comparable with the error of this approximation. Therefore different from $E_1$ as shown in Fig. \ref{delta1}, there is no tendency for $E_3$ to converge to zero as $n$ increases. From Fig. \ref{E3-3}, the small $E_3$ tends to be given by the small spin parameter $a$. This is caused by the fact that the small rotation speed of black hole corresponds to a small distorsion of photon ring's shape, which makes the approximation (\ref{deltar}) work better. Finally from Fig. \ref{E3-4}, one could find that the effect of the inclination angle $\theta_0$ is sub-dominated.

\section{Conclusions}
\label{sec:concl}

In this paper, we have proposed a method to test the Einstein equivalence principle of the electromagnetic law by observing the photon ring of black holes. Specifically, we start from a general Lagrangian (\ref{La}) that characterizes violation of the EEP. By applying the geometric optics approximation, we obtain the Eq. (\ref{system}) which implies the modified dispersion relation corresponding to the Lagrangian. In order to simplify and manifest the physical meaning of this system, we focus on the situation that the spacetime and the motion of photons have the spherical symmetry. This fact tells us that for the planar circular orbits, different polarized photons will sense different strength of gravitational potential and behaves as the violation of WEP as a result. The observable is expressed as Eq. (\ref{deltax}), which shows that the difference in the photon ring's size presented by two different linear polarized photons is proportionally connected to the difference of the corresponding EEP violation parameters. We also investigate the extend of the EEP violation to which the expression (\ref{deltax}) applies and display a few cases of specific EEP violation models.

For rotating black holes, the discussions would become more complicated. Our strategy is to select a representative model (\ref{metric}) to characterize the effects of black hole rotation. We compare the outcomes of the approximated expression (\ref{deltax}) having no rotation with those of the approximated expression (\ref{deltar}) having rotation. The results show that the large deviation ratio only occurs when the rotation of black hole is fast and the inclination angle of rotation axis is nearly edge-on or face-on. Similar to the discussions without rotation, we also estimate the accuracy of the expression (\ref{deltar}) by numerically solving the system, which is good for small magnitude of the EEP violation and small rotation speed of black holes.

In order to make the method in this paper workable, we need have the ability to distinguish photon ring in the photos of the supermassive black hole, which cannot be achieved with current observational capabilities \cite{Gralla:2020pra}. Recently, Johnson et al. show that the circular photon ring would manifest itself as a periodic visibility function on long interferometric baselines \cite{Johnson:2019ljv}, which thus makes the photon ring become distinct in the accretion background and the related lensing background. This work was subsequently extended to any shape of the photon ring by \cite{Gralla:2020nwp, Gralla:2020yvo} and the corresponding polarimetric signatures on long interferometric baselines by \cite{Himwich:2020msm}. The study of \cite{Gralla:2020srx} made the forecast that a space-based interferometry experiment can reach a accuracy level where the photon ring is remarkably insensitive to the astronomical source profile and can therefore be used to precisely test gravity. Furthermore, besides the appearance, a recent work also shows that the two-point correlation function of intensity fluctuations on the photon ring could also become an observable of the physics around black holes \cite{Hadar:2020fda}. In the near future, the next generation Event Horizon Telescope could have the ability to do the double band observations as well as the corresponding dual-polarizations \cite{Blackburn:2019bly,Haworth:2019urs}. All these theoretical and experimental advances provide us with opportunities to explore possible new physics in the strong gravitational field.

\section*{Acknowledgments}
We are grateful to Can-Min Deng, Damien Easson, Xin Ren, Sheng-Feng Yan, Ye-Fei Yuan and Pierre Zhang for stimulating discussions. This work is supported in part by the NSFC (Nos. 11653002, 11961131007, 11722327, 1201101448, 11421303), by the CAST (2016QNRC001), by the National Thousand Talents Program of China, by the Fundamental Research Funds for Central Universities, and by the USTC Fellowship for international cooperation. All numerics were operated on the computer clusters {\it LINDA \& JUDY} in the particle cosmology group at USTC. \\

\end{document}